\documentclass[runningheads]{llncs}
\usepackage{booktabs}
\usepackage{graphicx}
\usepackage{url}
\usepackage[T1]{fontenc}
\usepackage{multirow}
\usepackage{graphicx,verbatim}
\usepackage{color}
\usepackage{hyperref}

\begin{document}
\title{Evaluation of 3D Counterfactual Brain MRI Generation}

\author{Pengwei Sun\inst{1,2} \and
Wei Peng\inst{3} \and
Lun Yu Li\inst{4} \and
Yixin Wang\inst{5} \and
Kilian M. Pohl\inst{1,3,6}
}
%
\authorrunning{P. Sun et al.}
%
\institute{
Wu Tsai Neurosciences Institute, Stanford University, Stanford, CA, USA \and
Department of Radiology, Stanford University, Stanford, CA, USA \and
Dept. of Psychiatry \& Behavioral Sciences, Stanford University, Stanford, CA, USA \and
Department of Computer Science, Stanford University, Stanford, CA, USA \and
Department of Bioengineering, Stanford University, Stanford, CA, USA \and
Department of Electrical Engineering, Stanford University, Stanford, CA, USA
\email{kpohl@stanford.edu}
}

\maketitle              
\begin{abstract}
Counterfactual generation offers a principled framework for simulating hypothetical changes in medical imaging, with potential applications in understanding disease mechanisms and generating physiologically plausible data. However, generating realistic structural 3D brain MRIs that respect anatomical and causal constraints remains challenging due to data scarcity, structural complexity, and the lack of standardized evaluation protocols. In this work, we convert six generative models into 3D counterfactual approaches by incorporating an anatomy-guided framework based on a causal graph, in which regional brain volumes serve as direct conditioning inputs. Each model is evaluated with respect to composition, reversibility, realism, effectiveness and minimality on T1-weighted brain MRIs (T1w MRIs) from the Alzheimer's Disease Neuroimaging Initiative (ADNI). In addition, we test the generalizability of each model with respect to T1w MRIs of the National Consortium on Alcohol and Neurodevelopment in Adolescence (NCANDA).
Our results indicate that anatomically grounded conditioning successfully modifies the targeted anatomical regions; however, it exhibits limitations in preserving non-targeted structures. Beyond laying the groundwork for more interpretable and clinically relevant generative modeling of brain MRIs, this benchmark highlights the need for novel architectures that more accurately capture anatomical interdependencies.
Code: \url{https://github.com/pengwei2000/counterfactual_3DMRI}
\end{abstract}

\section{Introduction}
Deep generative models have demonstrated significant promise in medical imaging, enabling data-driven understanding and synthesis of complex anatomy captured by structural brain MRIs \cite{friedrich2024deepgenerativemodels3d}.
Counterfactual generation offers a principled way to explore hypothetical scenarios such as simulating anatomical changes due to age or disease by modeling causal relationships between metadata, anatomy, and image appearance \cite{melistas2024benchmarking}.
It performs interventions, direct manipulations of specific variables, to generate plausible alternatives to observed images.
Counterfactual generation holds significant promise in the medical domain, either by providing more realistic data to alleviate the problem of data scarcity in training deep learning models, or being used to explore biomarkers when the disease is successfully modeled for a specific subject \cite{gu2023biomedjourneycounterfactualbiomedicalimage}.

While most counterfactual approaches have been trained on natural images, the generation of 3D medical imaging presents unique challenges: anatomical features exhibit fine-grained variations while adhering to strict morphological constraints; data are often limited due to acquisition cost and privacy; and interventions must be anatomically and causally plausible \cite{Atad_2024}. To ensure adherence to causal constraints, one can use causal graphs, which explicitly encode the relationships between variables within conditional generative models \cite{Pearl_2009}. In such graphs, nodes represent variables and edges denote potential causal influences, allowing interventions on parent attributes propagating changes to their descendant nodes. In the context of brain MRI, prior works have constructed causal graphs in various ways \cite{melistas2024benchmarking,hvaemri,abdulaal2022deep,peng2024latent}, complicating direct cross-model comparisons. Among these studies, anatomical characterization is typically restricted to coarse metrics (such as total brain and ventricular volumes) thereby neglecting finer-grained structures, whose boundaries are not readily discernible on conventional structural MRI (such as  cingulate cortex). While some recent efforts have incorporated the volume of cortical structures into causal graphs \cite{peng2024latent}, systematic evaluation of interventions on these attributes remains lacking.

Evaluating counterfactuals of MRIs is challenging as ground-truth comparisons are generally unavailable. In addition, standard metrics (such as composition and realism) often overlook the anatomical consistency required for clinically meaningful synthesis. Furthermore, the accuracy of the model can vary across cohorts, which may arise from differences in hardware, acquisition protocols, annotation standards, and demographic or pathological distributions of patients. Deep generative models are powerful in memorizing the training set, facing the risk of overfitting when over-tuning hyperparameters on the validation set. As a result, models trained on data from a single cohort may inadvertently learn to rely on site-specific artifacts or biases that do not generalize well to unseen cohorts, which calls for a measurement of generalizability.
Finally, existing benchmarks are typically limited to 2D natural images or 2D slices of 3D medical images, failing to capture the volumetric and structural richness of clinical data \cite{monteiro2023measuringaxiomaticsoundnesscounterfactual,melistas2024benchmarking}.

To address the gap in fine-grained anatomical intervention, we adapt 6 deep generative models for anatomy-guided counterfactual generation, where image synthesis is explicitly conditioned on the volume of cortical regions and the ventricles. This localized intervention approach aims to modify only the targeted region while preserving the rest of the brain structure, aligning closely with clinical interpretability and biological plausibility.
To standardize evaluation metrics, we propose a unified framework for benchmarking 3D counterfactual generation of brain MRI. It consists of recording measures typically from 2D natural image benchmarks \cite{melistas2024benchmarking} (i.e., composition, realism, effectiveness, minimality) evaluated on ADNI dataset. To address the complexity associated with T1w MRIs, we measure reversibility (the ability to recover the original image after a sequence of forward and reverse interventions) on ADNI and generalizability on the NCANDA dataset. By doing so, we hope to provide a framework for the standardized evaluation and fair comparison of counterfactual approaches of 3D brain MRIs, ultimately fostering the development of more reliable and clinically relevant generative models for medical imaging.

\section{Methods}
\begin{figure}[t]
\centering
\includegraphics[width=4cm]{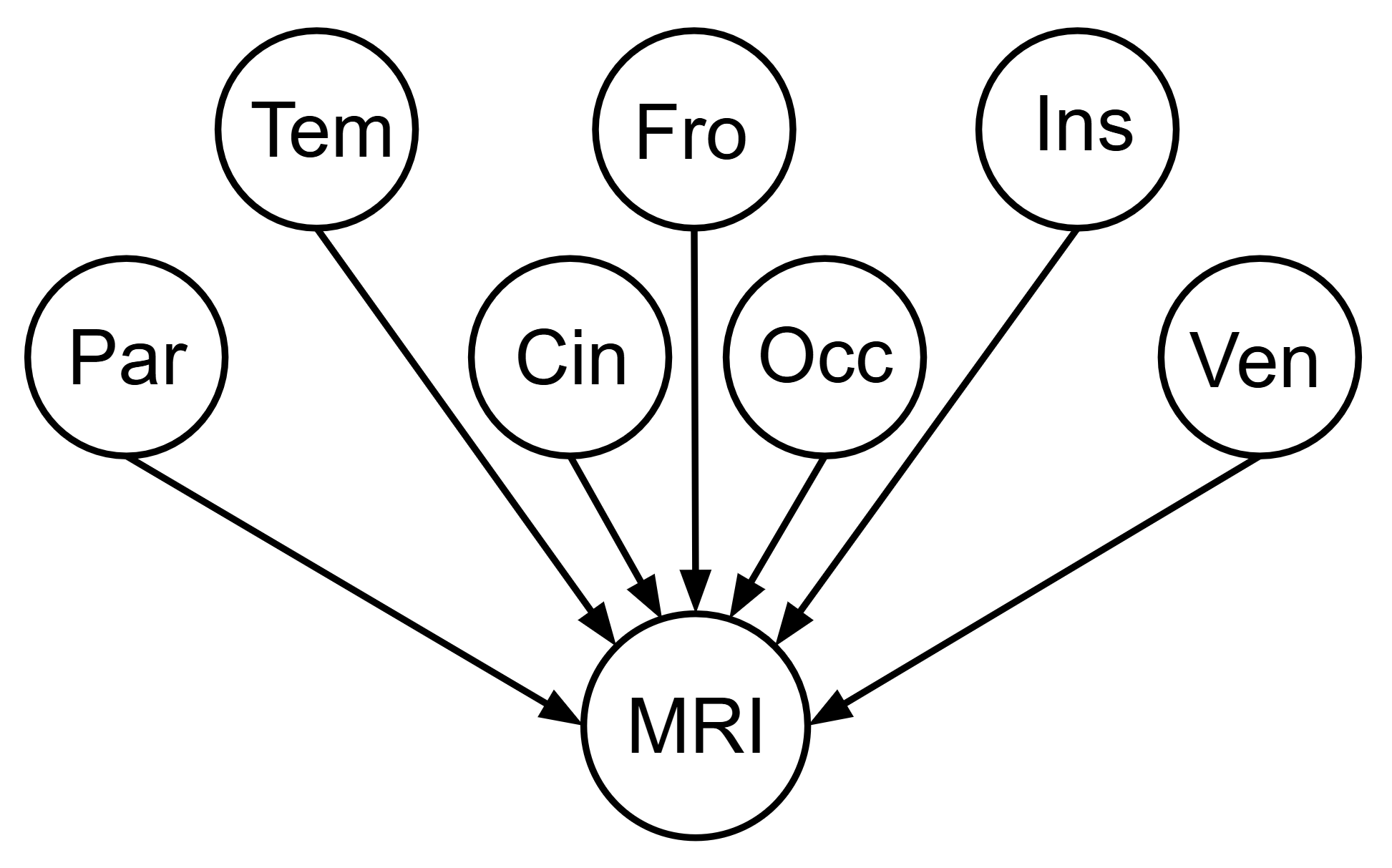}
\caption{A causal graph for 3D brain MRIs.} \label{fig1}
\end{figure}

\noindent \textbf{\underline{Causal Graphs}} We build a causal graph based on the regional volume measures extracted from t1w brain MRIs (Fig.~\ref{fig1}). Specifically, we include the volumes of ventricle (\textsf{Ven}), and several brain regions: the parietal lobe (\textsf{Par}), cingulate cortex (\textsf{Cin}), occipital lobe (\textsf{Occ}), temporal lobe (\textsf{Tem}), frontal lobe (\textsf{Fro}) and insula (\textsf{Ins}). The edges represent causality between the anatomical volumes and the MRI scan (\textsf{MRI}).

\noindent \textbf{\underline{Models}} For counterfactual generation of 3D MRI (\textsf{MRI}), we include 6 models: three Variational Autoencoder (VAE; basic version \cite{vae}, Hierarchical Variational Autoencoder (HVAE) \cite{hvae} and VAE with Generalized Linear Model (VAE-GLM)~\cite{peng2024latent}) and three Generative Adversarial Networks (GAN; basic version\cite{gan}, a finetuned GAN, and Hierarchical Amortized GAN (HA-GAN) \cite{hagan}).

Specifically, VAE is adapted from a 2D VAE architecture \cite{melistas2024benchmarking}, where 
the latent variables and conditioned parent attributes are passed through a decoder to reconstruct counterfactuals. The conditional HVAE model is implemented following \cite{hvaemri}. The latent variables are amortized and injected sequentially into the encoder and decoder. Instead of finding latent distributions using down-sampled images in VAE, HVAE maps the latent variables at multiple intermediate image resolutions. VAE-GLM utilizes a Generalized Linear Model (GLM)~\cite{peng2024metadata} to integrate metadata into the generation process. Here we specify the metadata to be volumes of the 7 brain regions. 


With respect to the conditional GANs, they include an encoder to produce the latent representations given images and parent attributes. We extend the 2D experimental settings in \cite{XIA2021102169} and use Fourier embeddings \cite{10.3389/fcvm.2022.983091} to encode the parent attributes. To improve the reconstruction quality, we use the cyclic cost minimization approach \cite{DOGAN2020338} to finetune the encoder, referred to as GAN-Finetuned. We incorporate HA-GAN for conditional generation by first producing a cropped, low-resolution image that is subsequently up-sampled to full-resolution, which effectively reduces memory constraints and patchy artifacts commonly observed in traditional 3D GANs. 

\noindent \textbf{\underline{Datasets}} T1w brain MRIs were obtained from the Alzheimer’s Disease Neuroimaging Initiative (ADNI) \cite{doi:10.1212/WNL.0b013e3181cb3e25} database (Data Release 1, 2, 3 and GO). There are 4578 scans from 1511 subjects passed through the preprocessing pipeline \cite{10.1007/978-3-031-43993-3_2,ZHAO2021102051}, including denoising, homogeneity-correction, skull-stripping, intensity normalization, and affine alignment to a template, which resulted in the 1 mm isotropic MRIs consisting of 144 × 176 × 144 voxels.
The age of subjects ranges from 55.25 to 96.00. 55\% of subjects are male. 
We randomly split the dataset into training (90\%) and test (10\%) sets based on the subject id.

To assess generalizability, we randomly sampled 582 MRI scans of 106 subjects from the NCANDA\_PUBLIC\_6Y\_STRUCTURAL\_V01 data release of the National Consortium on Alcohol and NeuroDevelopment in Adolescence (NCANDA) \cite{doi:10.15288/jsad.2015.76.895}. All scans underwent the same preprocessing pipeline as applied to the ADNI dataset. 
The age of subjects ranges from 12.00 to 28.02. 49\% of subjects are male.

\noindent \textbf{\underline{Evaluation Metrics}} The six categories of evaluation metrics are defined next with the first four being inspired by \cite{melistas2024benchmarking} and reversibility and generalizability being introduced in order to account for the  complexity associated with T1w MRIs:

\textbf{Composition} Null intervention directly passes the latent distribution to the generator without interventions on parent attributes. Under passes of null intervention, the composition measures image stability and consistency. A high composition score indicates the model's ability to preserve the unique anatomy of the subject in the medical domain. Discrepancies between the original and generated images are captured with $l_1$ distance and Structural Similarity Index Measure (SSIM) \cite{1284395}.
Following established conventions in counterfactual evaluation \cite{melistas2024benchmarking}, we measure composition after both 1 and 10 consecutive passes through the model.

\textbf{Effectiveness and Minimality} The effectiveness quantifies how accurately a model modifies the targeted attribute (e.g., the volume of the ventricles), while minimality records how the other non-intervened variables (i.e., volumes of the other brain regions) have changed. To measure effectiveness and minimality, we view SynthSeg~\cite{billot_robust_2023} as an oracle to conduct the segmentation and cortex parcellation on the counterfactuals. After intervening on one attribute, SynthSeg determines the volumes of cortical areas and ventricle of the counterfactuals. For each region, the volume scores across all subjects are normalized between -1 and 1.

\textbf{Realism} Realism evaluates the similarity between counterfactuals and real MRIs. We employ the Fréchet Inception Distance (FID) \cite{heusel2018ganstrainedtimescaleupdate} to quantify image fidelity by comparing the distributions of real and counterfactual images. MedicalNet \cite{chen2019med3d}, a 3D ResNet pretrained on large medical datasets, is used as the domain-specific feature extractor to calculate FID scores. 

\textbf{Reversibility} measures the robustness of a model under cycles of reverse intervention. The target attribute is conditioned to some random value in the first pass of a reversibility cycle, then conditioned back to its original value in the second pass of a cycle. We evaluate reversibility at both 1 and 3 cycles to observe both immediate and cumulative effects of sequential interventions. Multiple cycles expose compounding errors and drift that might remain imperceptible in shorter sequences of intervention. Models with low reversibility scores, calculated using $l_1$ distance in the 3D image space, are more robust and consistent to interventions.

\textbf{Generalizability} 
To evaluate the generalizability, we train the deep generative models on ADNI dataset, and evaluate the generative ability on the NCANDA dataset. Similar to effectiveness, we intervene on one anatomical attribute and evaluate the MAE between the target and counterfactual volumes of the intervened brain region. Lower MAE signifies better generalization, implying better adaptation to unseen data. 

\begin{table}[t]
\centering
\caption {\label{tab:composition} Composition, reversibility, and realism scores on ADNI dataset. Top scores are in bold. An upward arrow indicates that larger values are interpreted as better results.} 
\resizebox{10cm}{!}{
\begin{tabular}{c|ccccc|cc|c}
\hline
 & \multicolumn{5}{c|}{\textbf{Composition}} &\multicolumn{2}{c|}{\textbf{Reversibility}} & \textbf{Realism}\\
 \textbf{Model} & \multicolumn{2}{c}{$l_1$ \textbf{distance} $\downarrow$}  & & \multicolumn{2}{c|}{\textbf{SSIM} $\uparrow$} &   \multicolumn{2}{c|}{$l_1$ \textbf{distance} $\downarrow$}&\textbf{FID} $\downarrow$\\
\cline{2-4} \cline{5-6} \cline{7-8} 
        &  1 pass  & 10 passes    &           
	&  1 pass & 10 passes   & 1 cycle & 3 cycles                            \\
\hline
VAE   &   0.115   &  0.143   &  &0.750 &0.680 &    0.130  & 0.141&9.086  \\
HVAE    & \textbf{0.000} &\textbf{0.000}&&\textbf{1.000}&\textbf{1.000}&    \textbf{0.003}  &    \textbf{0.007}&\textbf{0.026} \\
VAE-GLM  & 0.030 & 0.097&&0.908&0.633  &   0.038  & 0.069&0.311\\
GAN      &0.245&0.409&&0.293&0.063 &  0.287    &  0.363&70.853 \\
GAN-Finetuned &0.435 &0.596&&0.283&0.044& 0.417   & 0.542&51.868\\
HA-GAN &0.175&0.175&&0.551&0.552 &   0.175  & 0.176 &8.147\\
\hline
\end{tabular}}
\end{table}

%
%
\section{Results}
\noindent \textbf{Composition} 
Table \ref{tab:composition} demonstrates the exceptional composition performance of HVAE in both distance metrics, maintaining high image fidelity even after 10 passes. 
VAE-GLM has relatively low $l_1$ distance and high SSIM after one pass but worsens after ten passes. GAN-based models have the worst composition scores after one pass and further deteriorate after ten passes. Supplementary Fig. \ref{fig3} confirms these quantitative results of composition. VAE counterfactuals are blurry, losing fine anatomical structures of brain MRIs. This issue comes from the difficulty to reconstruct the high-resolution images directly from latent space. GAN-based models generate unrealistic images and the brightness of brains is prone to change. 

\begin{figure}
\centering
\includegraphics[width=\textwidth]{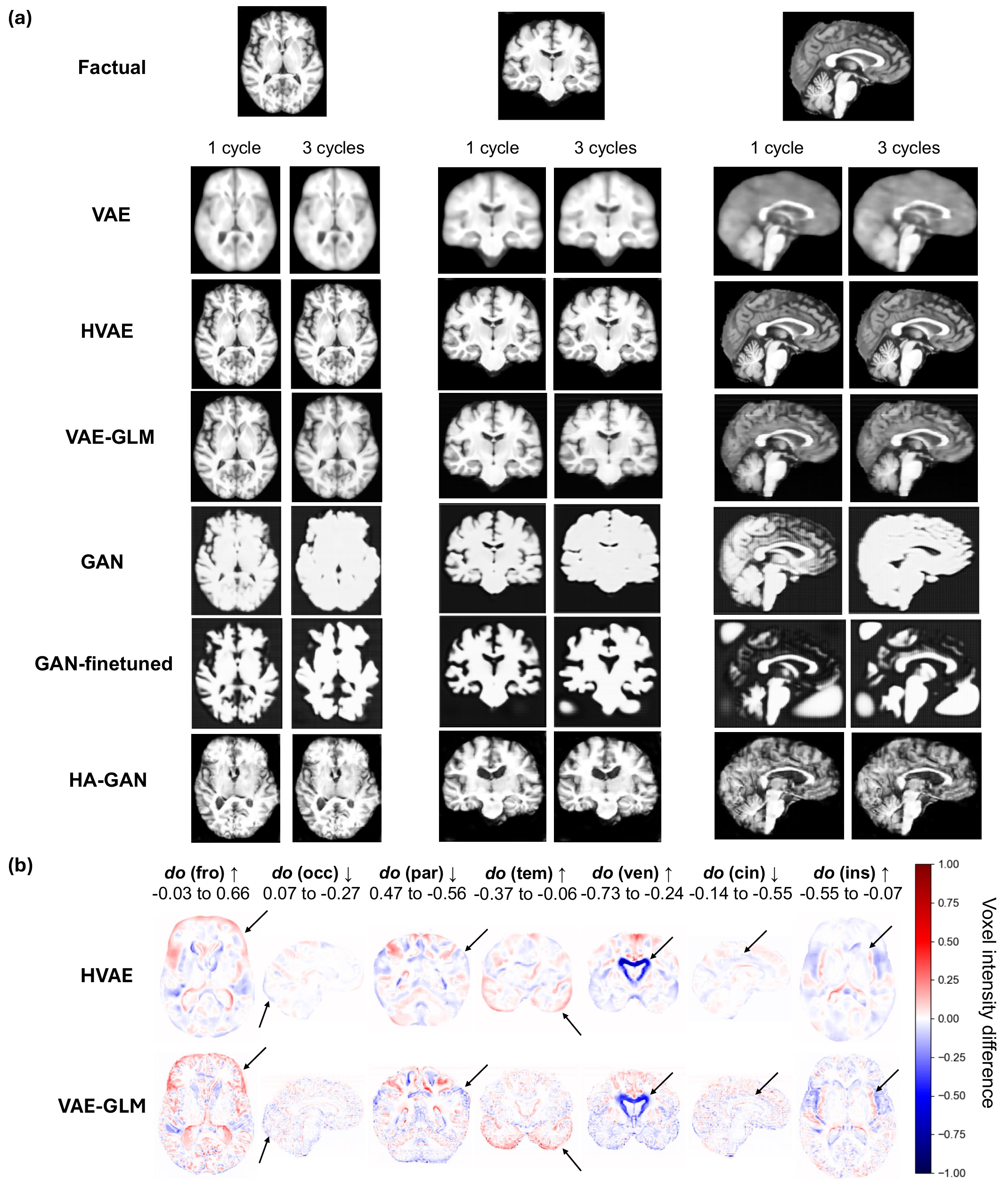}
\caption{(a) Qualitative evaluation of reversibility on ADNI dataset. The factual image, counterfactual image after 1, 3 cycles are shown for each model. (b) Difference between factual images and counterfactuals. Each column represents an intervention on the volume of one anatomical structure. Bold arrows in the images indicate relevant anatomical volume changes. The original values and target values are shown above the counterfactuals. An upward arrow after \textit{do}($\cdot$) indicates that the volume of that region was increased during intervention.} \label{effectiveness}
\end{figure}

\noindent \textbf{Reversibility}
We apply intervention cycles on one attribute at a time and report averaged reversibility scores across all attributes. After one or three cycles of reverse intervention, HVAE generates counterfactuals that are closest to the factual images, followed by VAE-GLM (Table~\ref{tab:composition}). The qualitative results are shown in Fig. \ref{effectiveness}a. GAN-based models produce counterfactuals farther from the original images, lacking the ability to preserve anatomical brain structures after three cycles. HA-GAN and VAE generate brain outlines, but HA-GAN enlarges ventricles after cycles and VAE generates blurry images.

\noindent \textbf{Realism}
The factual and counterfactual images share a similar distribution when the FID score is lower. In other words, the counterfactuals are more "real".
HVAE has the lowest FID scores, followed by VAE-GLM (Table~\ref{tab:composition}). GAN-based models tend to produce unrealistic images that deviate from the original image distribution, though HA-GAN offers some alleviation.

\noindent \textbf{Effectiveness and Minimality}
Table \ref{tab:effectiveness} lists the mean absolute error (MAE) with respect to interventions on the cortical volume scores. According to this table, VAE consistently achieves the lowest mean absolute error (MAE) for all targeted intervention, such as parietal lobe volume MAE when applying an intervention on the volume of the parietal lobe (\textit{do}(\textsf{Par})), indicating it is the most effective model for anatomical modifications. HA-GAN and VAE-GLM also demonstrate strong effectiveness across most interventions, with the exception of \textit{do}(\textsf{Ven}).

\begin{table}[t]
\centering
\caption {\label{tab:effectiveness} Effectiveness scores on ADNI dataset and generalizability scores on NCANDA dataset. For example, \textit{do}(\textsf{Fro}) denotes an intervention on the frontal lobe volume, i.e., the model is conditioned to modify only the frontal lobe volume while preserving other attributes.}
\resizebox{12cm}{!}{
\begin{tabular}{c|c|c|ccccccc}
\hline
 \textbf{Metrics}& \textbf{Dataset}&\textbf{Model} &   \multicolumn{7}{c}{\textbf{Intervening target volume MAE} $\downarrow$}    \\ 
\cline{4-10} 
       && &  \textit{do}(\textsf{Fro})  & \textit{do}(\textsf{Par})  &\textit{do}(\textsf{Tem})&\textit{do}(\textsf{Occ}) 
        &\textit{do}(\textsf{Cin})	& \textit{do}(\textsf{Ins})&\textit{do}(\textsf{Ven})                               \\
\hline
&&VAE&\textbf{0.028}&\textbf{0.033}&\textbf{0.016}&\textbf{0.041}&\textbf{0.032}&\textbf{0.018}&\textbf{0.028}\\
&&HVAE   
&0.065&0.067&0.052&0.073&0.060&0.046&0.082 \\
Effectiveness&ADNI&VAE-GLM &0.055&0.061&0.044&0.068&0.055&0.041&0.104 \\
&&GAN      
&0.077&0.074&0.064&0.082&0.061&0.058&0.189\\
&&GAN-Finetuned 
&0.119&0.139&0.082&0.103&0.096&0.076&0.198\\
&&HA-GAN 
&0.041&0.051&0.040&0.055&0.044&0.035&0.112 \\
\hline
\hline
&&VAE    &0.144&0.148&0.052&\textbf{0.056}&0.101&\textbf{0.017}&0.488\\
&&HVAE   &0.053&0.056&0.046&0.079&0.053&0.045&\textbf{0.082} \\
Generalizability&NCANDA&VAE-GLM &\textbf{0.042}&\textbf{0.049}&\textbf{0.027}&0.072&\textbf{0.045}&0.039&0.096 \\
&&GAN      &0.067&0.063&0.065&0.087&0.086&0.069&0.140\\
&&GAN-Finetuned &0.137&0.148&0.112&0.099&0.144&0.113&0.116\\
&&HA-GAN &0.104&0.151&0.080&0.064&0.075&0.037&0.443 \\
\hline
\end{tabular}}
\end{table}

However, our minimality analysis (Supplementary Table \ref{tab:minimality}) reveals a concerning pattern across all models: the MAE for non-intervened attributes is consistently higher than the effectiveness scores for \textsf{Fro}, \textsf{Cin}, \textsf{Occ}, and \textsf{Ins}. Intervening on \textsf{Par} or \textsf{Tem} results in MAEs comparable to effectiveness scores for \textsf{Tem} and \textsf{Par}, respectively, but intervening on other regions leads to larger MAEs of \textsf{Tem} and \textsf{Par}. This indicates that while models can effectively modify targeted attributes, they struggle to preserve other anatomical features during intervention. Compared to VAE, GAN-based methods generally received better minimality scores.
However, the overall preservation of non-intervened volumes remains inadequate across all tested approaches.

This observation is also illustrated in Fig. \ref{effectiveness}b, which shows the impact of the intervention on a single anatomical attribute by visualizing the difference between the factual images and counterfactuals. Interventions conditioned on ventricles (\textsf{Ven}) and temporal lobes (\textsf{Tem}) result in an evident increase in their volumes. There is also a noticeable volume decrease in cingulate (\textsf{Cin}) and occipital lobes (\textsf{Occ}). \textit{do}(\textsf{Fro}) significantly expands frontal volume in counterfactuals, accompanied by minor, potentially spurious, volume changes in parietal lobes (\textsf{Par}) and ventricles. \textit{do}(\textsf{Par}) induces parietal volume decrease coupled with frontal volume increase. \textit{do}(\textsf{Ins}) is also effective despite shrunken neighboring cortex. The counterfactual images and difference plots of all other models are shown in Supplementary Fig. \ref{fig4}.

\noindent \textbf{Generalizability}
on NCANDA, VAE-GLM performs best when evaluating the intervention on \textsf{Fro}, \textsf{Par}, and \textsf{Tem} (Table \ref{tab:effectiveness}). VAE has the lowest MAE in terms of \textsf{Occ} and \textsf{Ins}. And HVAE does better with \textsf{Ven} modification. Although not achieving top effectiveness scores on ADNI dataset, VAE-GLM generalizes well to the unseen dataset.

\section{Discussion}
Our evaluation across six categories of metrics - composition, reversibility, realism, effectiveness, minimality, and generalizability - reveals distinct performance trade-offs among the models.
Autoencoder-based approaches more readily incorporated the intended anatomical modifications, yielding high effectiveness, whereas adversarially trained models (GAN variants) demonstrated slightly better preservation of non-targeted features, though still far from ideal.
The VAE achieved the most accurate targeted changes (highest effectiveness), but its counterfactual outputs were noticeably blurred and less anatomically faithful, indicating poorer composition stability and realism.
HVAE and VAE-GLM achieved the best balance across all metrics except for minimality. Although their intervention accuracy was slightly lower than VAE's, these models excelled in preserving overall brain anatomy and image quality, exhibiting superior composition consistency and lower drift in reversibility cycles. HVAE's hierarchical multi-resolution latent architecture allows it to capture fine anatomical details while still responding accurately to interventions. VAE-GLM's explicit disentanglement of each attribute's effect within the latent space enables more controlled modifications.
Moreover, VAE-GLM demonstrated outstanding generalizability: it achieved the lowest MAE on most anatomical interventions in an external test dataset. HVAE also generalized well in certain scenarios (e.g., ventricle volume changes). 
GAN-based methods produced sharp images but struggled with counterfactual fidelity. Interventions often led to patchy or unrealistic details, resulting in low realism and degraded reversibility over repeated edits.

However, the low performance with respect to minimality of all models points to a fundamental limitation in current approaches to counterfactual generation. Despite achieving reasonable effectiveness in modifying targeted attributes, the models frequently induce unintended changes in other anatomical regions, suggesting inadequate disentanglement of causal effects in the latent space. This challenge likely stems from the complex interdependencies between anatomical structures that are inadequately modeled in current causal frameworks. New architectures specifically designed to better enforce minimality constraints are urgently needed for medically plausible counterfactual generation.

\section{Conclusion}
We present a unified benchmarking framework for 3D counterfactual generation of brain MRI, addressing a critical need for standardized evaluation in medical generative modeling. By conditioning on the volumes of specific brain regions captured in a causal graph, our framework enables interpretable interventions. We assess each model across six axes: composition, reversibility, realism, effectiveness, minimality, and generalizability.
This assessment revealed a significant limitation that these models can effectively modify targeted anatomical regions but fail to preserve non-intervened structures. This finding underscores the need for new counterfactual approaches that are designed to accurately capture the constraints across anatomical brain structures. We hope that this gap in current technology will catalyze further research on interpretable, reliable counterfactual generation and facilitate its integration into downstream applications such as biomarker discovery and modeling of disease progression.

\begin{credits}
\subsubsection{\ackname} This work was partly supported by the Wu Tsai Neurosciences Institute, the National Institute of Health (AA021697, DA057567), and by the Stanford University Human-Centered Artificial Intelligence.

NCANDA data collection and distribution were supported by NIH funding AA021681, AA021690, AA021691, AA021692, AA021695, AA021696, AA021697. They are made publicly accessible via \url{https://nda.nih.gov/edit_collection.html?id=4513}.  

\subsubsection{\discintname}
The authors have no competing interests to declare that are
relevant to the content of this article.
\end{credits}

%
%
%
%
\newpage
\bibliographystyle{splncs04}
\bibliography{egbib}

\begin{thebibliography}{10}
\providecommand{\url}[1]{\texttt{#1}}
\providecommand{\urlprefix}{URL }
\providecommand{\doi}[1]{https://doi.org/#1}

\bibitem{abdulaal2022deep}
Abdulaal, A., Castro, D.C., Alexander, D.C.: Deep structural causal modelling of the clinical and radiological phenotype of alzheimer{\textquoteright}s disease. In: NeurIPS 2022 Workshop on Causality for Real-world Impact (2022)

\bibitem{Atad_2024}
Atad, M., Schinz, D., Moeller, H., Graf, R., Wiestler, B., Rueckert, D., Navab, N., Kirschke, J.S., Keicher, M.: Counterfactual explanations for medical image classification and regression using diffusion autoencoder. Machine Learning for Biomedical Imaging  \textbf{2}(iMIMIC 2023),  2103–2125 (Sep 2024). \doi{10.59275/j.melba.2024-4862}

\bibitem{billot_robust_2023}
Billot, B., Colin, Magdamo~Cheng, Y., Das, S., Iglesias, J.E.: {Robust} machine learning segmentation for large-scale analysis of heterogeneous clinical brain {MRI} datasets. {Proceedings} of the {National} {Academy} of {Sciences} ({PNAS})  \textbf{120}(9),  1--10 (2023). \doi{10.1073/pnas.2216399120}

\bibitem{doi:10.15288/jsad.2015.76.895}
Brown, S.A., Brumback, T., Tomlinson, K., Cummins, K., Thompson, W.K., Nagel, B.J., De~Bellis, M.D., Hooper, S.R., Clark, D.B., Chung, T., Hasler, B.P., Colrain, I.M., Baker, F.C., Prouty, D., Pfefferbaum, A., Sullivan, E.V., Pohl, K.M., Rohlfing, T., Nichols, B.N., Chu, W., Tapert, S.F.: The national consortium on alcohol and neurodevelopment in adolescence ({NCANDA}): A multisite study of adolescent development and substance use. Journal of Studies on Alcohol and Drugs  \textbf{76}(6),  895--908 (2015). \doi{10.15288/jsad.2015.76.895}, pMID: 26562597

\bibitem{10.3389/fcvm.2022.983091}
Campello, V.M., Xia, T., Liu, X., Sanchez, P., Martín-Isla, C., Petersen, S.E., Seguí, S., Tsaftaris, S.A., Lekadir, K.: Cardiac aging synthesis from cross-sectional data with conditional generative adversarial networks. Frontiers in Cardiovascular Medicine  \textbf{9} (2022). \doi{10.3389/fcvm.2022.983091}

\bibitem{chen2019med3d}
Chen, S., Ma, K., Zheng, Y.: {Med3D}: Transfer learning for {3D} medical image analysis. arXiv preprint arXiv:1904.00625  (2019)

\bibitem{hvaemri}
De~Sousa~Ribeiro, F., Xia, T., Monteiro, M., Pawlowski, N., Glocker, B.: High fidelity image counterfactuals with probabilistic causal models. In: Proceedings of the 40th International Conference on Machine Learning. Proceedings of Machine Learning Research, vol.~202, pp. 7390--7425 (23--29 Jul 2023), \url{https://proceedings.mlr.press/v202/de-sousa-ribeiro23a.html}

\bibitem{DOGAN2020338}
Dogan, Y., Keles, H.Y.: Semi-supervised image attribute editing using generative adversarial networks. Neurocomputing  \textbf{401},  338--352 (2020), \url{https://doi.org/10.1016/j.neucom.2020.03.071}

\bibitem{friedrich2024deepgenerativemodels3d}
Friedrich, P., Frisch, Y., Cattin, P.C.: Deep Generative Models for 3D Medical Image Synthesis, pp. 255--278. Springer Nature Switzerland, Cham (2025), \url{https://doi.org/10.1007/978-3-031-80965-1_13}

\bibitem{gan}
Goodfellow, I.J., Pouget-Abadie, J., Mirza, M., Xu, B., Warde-Farley, D., Ozair, S., Courville, A., Bengio, Y.: Generative adversarial nets. In: Advances in Neural Information Processing Systems. vol.~27. Curran Associates, Inc. (2014), \url{https://proceedings.neurips.cc/paper_files/paper/2014/file/f033ed80deb0234979a61f95710dbe25-Paper.pdf}

\bibitem{gu2023biomedjourneycounterfactualbiomedicalimage}
Gu, Y., Yang, J., Usuyama, N., Li, C., Zhang, S., Lungren, M.P., Gao, J., Poon, H.: Biomedjourney: Counterfactual biomedical image generation by instruction-learning from multimodal patient journeys (2023), \url{https://arxiv.org/abs/2310.10765}

\bibitem{heusel2018ganstrainedtimescaleupdate}
Heusel, M., Ramsauer, H., Unterthiner, T., Nessler, B., Hochreiter, S.: Gans trained by a two time-scale update rule converge to a local nash equilibrium. In: Advances in Neural Information Processing Systems. vol.~30. Curran Associates, Inc. (2017), \url{https://proceedings.neurips.cc/paper_files/paper/2017/file/8a1d694707eb0fefe65871369074926d-Paper.pdf}

\bibitem{vae}
Higgins, I., Matthey, L., Pal, A., Burgess, C.P., Glorot, X., Botvinick, M.M., Mohamed, S., Lerchner, A.: {beta-VAE}: Learning basic visual concepts with a constrained variational framework. In: International Conference on Learning Representations (2016), \url{https://api.semanticscholar.org/CorpusID:46798026}

\bibitem{melistas2024benchmarking}
Melistas, T., Spyrou, N., Gkouti, N., Sanchez, P., Vlontzos, A., Panagakis, Y., Papanastasiou, G., Tsaftaris, S.A.: Benchmarking counterfactual image generation. In: The Thirty-eight Conference on Neural Information Processing Systems Datasets and Benchmarks Track (2024)

\bibitem{monteiro2023measuringaxiomaticsoundnesscounterfactual}
Monteiro, M., Ribeiro, F.D.S., Pawlowski, N., Castro, D.C., Glocker, B.: Measuring axiomatic soundness of counterfactual image models. In: The Eleventh International Conference on Learning Representations (2023)

\bibitem{Pearl_2009}
Pearl, J.: Causality. Cambridge University Press, 2 edn. (2009)

\bibitem{10.1007/978-3-031-43993-3_2}
Peng, W., Adeli, E., Bosschieter, T., Park, S.H., Zhao, Q., Pohl, K.M.: Generating realistic brain {MRIs} via a conditional diffusion probabilistic model. In: Medical Image Computing and Computer Assisted Intervention -- MICCAI 2023. pp. 14--24. Springer Nature Switzerland, Cham (2023)

\bibitem{peng2024metadata}
Peng, W., Bosschieter, T., Ouyang, J., Paul, R., Sullivan, E.V., Pfefferbaum, A., Adeli, E., Zhao, Q., Pohl, K.M.: Metadata-conditioned generative models to synthesize anatomically-plausible {3D} brain {MRIs}. Medical Image Analysis  \textbf{98},  103325 (2024)

\bibitem{peng2024latent}
Peng, W., Xia, T., Ribeiro, F.D.S., Bosschieter, T., Adeli, E., Zhao, Q., Glocker, B., Pohl, K.M.: Latent {3D} brain {MRI} counterfactual with structural causal model. Deep Generative Models for Medical Image Computing and Computer Assisted Intervention  (2025)

\bibitem{doi:10.1212/WNL.0b013e3181cb3e25}
Petersen, R.C., Aisen, P.S., Beckett, L.A., Donohue, M.C., Gamst, A.C., Harvey, D.J., Jack, C.R., Jagust, W.J., Shaw, L.M., Toga, A.W., Trojanowski, J.Q., Weiner, M.W.: Alzheimer's disease neuroimaging initiative ({ADNI}). Neurology  \textbf{74}(3),  201--209 (2010). \doi{10.1212/WNL.0b013e3181cb3e25}

\bibitem{hagan}
Sun, L., Chen, J., Xu, Y., Gong, M., Yu, K., Batmanghelich, K.: Hierarchical amortized {GAN} for {3D} high resolution medical image synthesis. IEEE Journal of Biomedical and Health Informatics  \textbf{26}(8),  3966--3975 (2022). \doi{10.1109/JBHI.2022.3172976}

\bibitem{hvae}
Sønderby, C.K., Raiko, T., Maaløe, L., Sønderby, S.K., Winther, O.: Ladder variational autoencoders. In: Advances in Neural Information Processing Systems. vol.~29. Curran Associates, Inc. (2016), \url{https://proceedings.neurips.cc/paper_files/paper/2016/file/6ae07dcb33ec3b7c814df797cbda0f87-Paper.pdf}

\bibitem{1284395}
Wang, Z., Bovik, A., Sheikh, H., Simoncelli, E.: Image quality assessment: from error visibility to structural similarity. IEEE Transactions on Image Processing  \textbf{13}(4),  600--612 (2004). \doi{10.1109/TIP.2003.819861}

\bibitem{XIA2021102169}
Xia, T., Chartsias, A., Wang, C., Tsaftaris, S.A.: Learning to synthesise the ageing brain without longitudinal data. Medical Image Analysis  \textbf{73},  102169 (2021), \url{https://doi.org/10.1016/j.media.2021.102169}

\bibitem{ZHAO2021102051}
Zhao, Q., Liu, Z., Adeli, E., Pohl, K.M.: Longitudinal self-supervised learning. Medical Image Analysis  \textbf{71},  102051 (2021), \url{https://doi.org/10.1016/j.media.2021.102051}

\end{thebibliography}

\newpage
\section{Additional results}
\begin{figure}
\centering
\includegraphics[width=\linewidth]{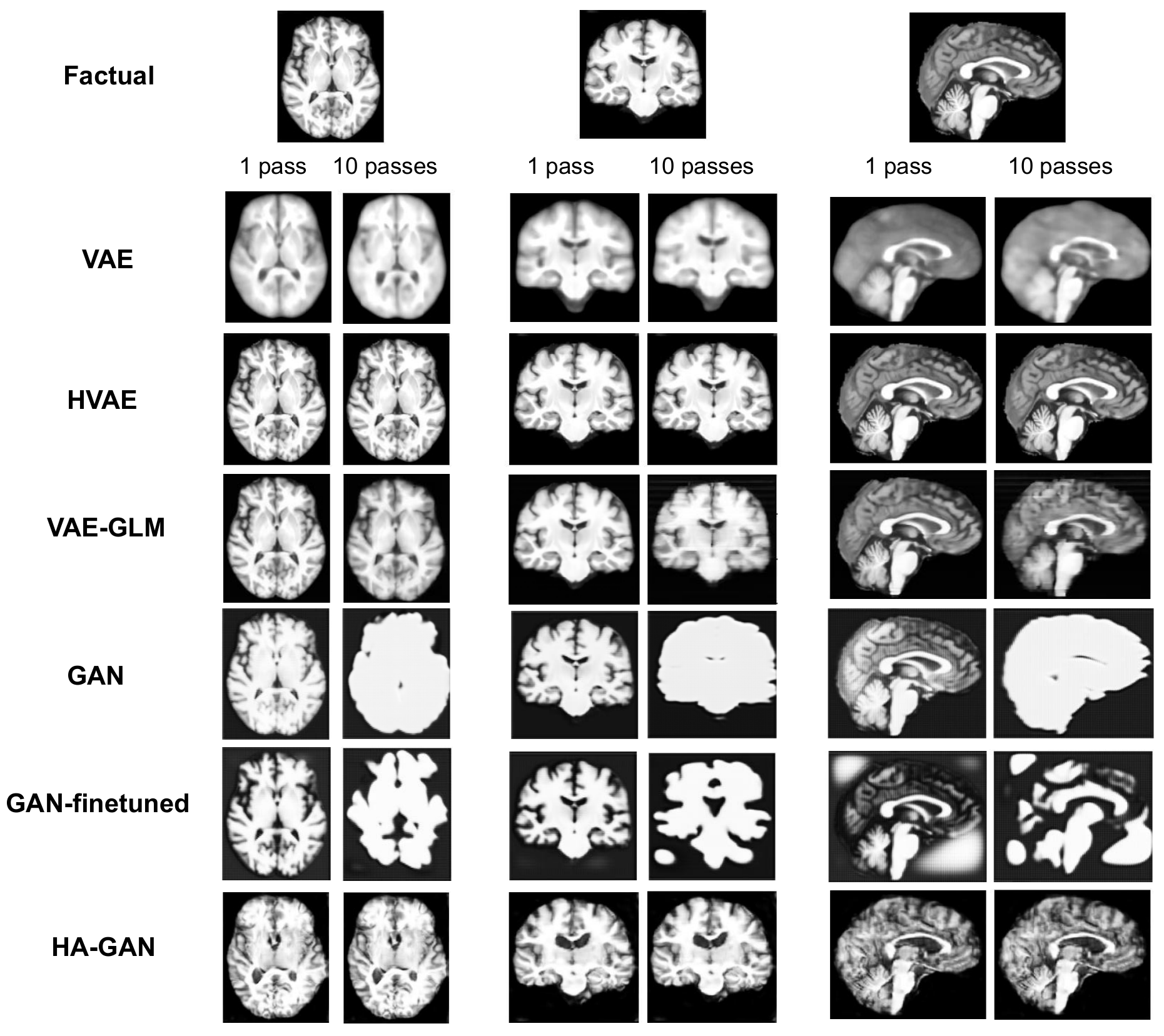}
\caption{Qualitative evaluation of composition on ADNI dataset. The factual image, counterfactual image after 1, 10 passes are shown for each model.} \label{fig3}
\end{figure}

\begin{figure}[t]
\centering
\includegraphics[width=0.8\linewidth]{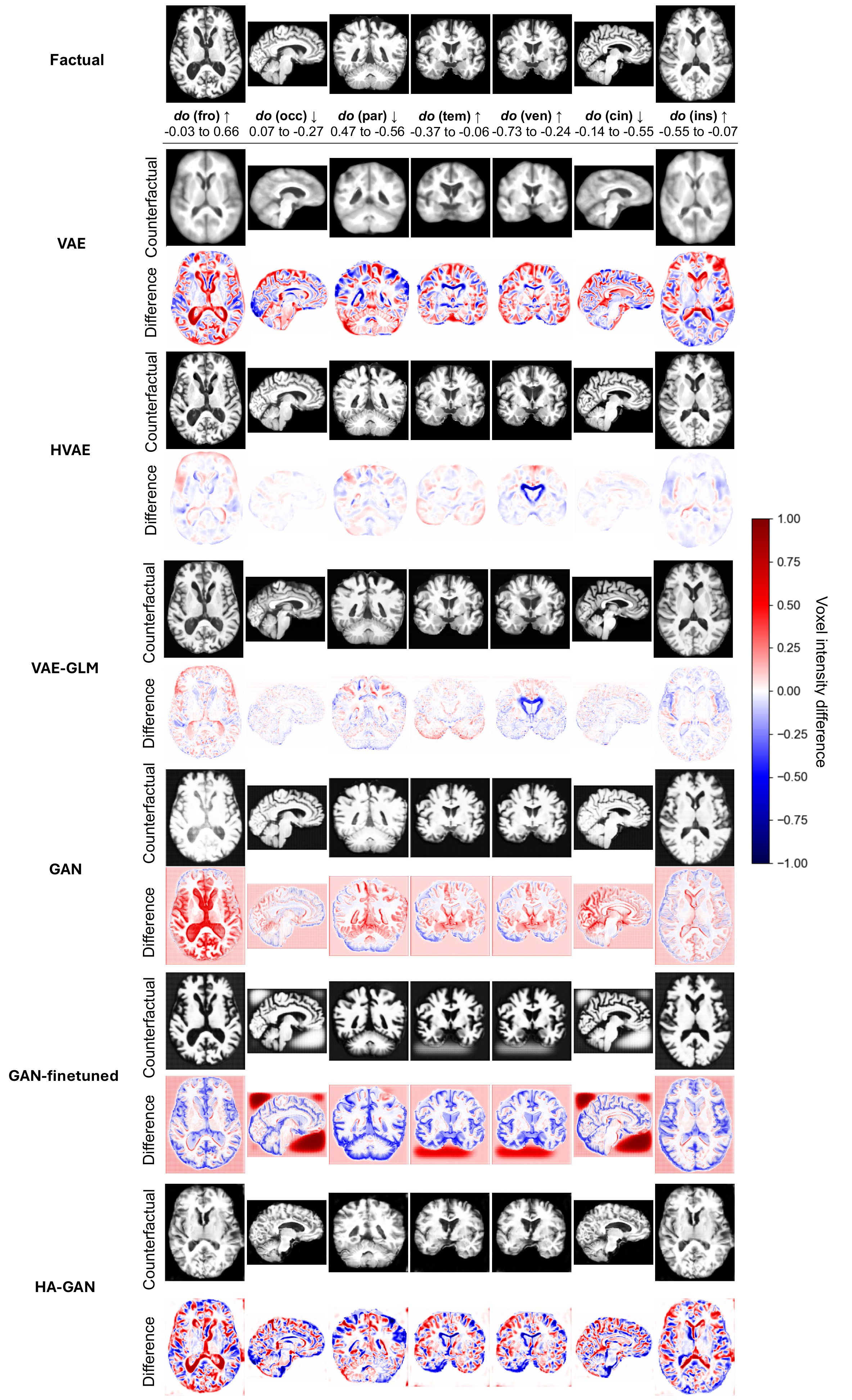}
\caption{Qualitative evaluation of effectiveness on ADNI dataset. The factual image, counterfactual image, and their difference are shown for each model. The original values and target values are shown below the factual images. An upward arrow after \textit{do}($\cdot$) indicates that the volume of that region was increased during intervention.}
\label{fig4}
\end{figure}

\begin{table}[t]
\centering
\caption {\label{tab:minimality} Minimality scores on ADNI dataset. An downward arrow indicates that smaller values are interpreted as better results.} 
\resizebox{10cm}{!}{
\begin{tabular}{c|ccccccc}
\hline
 \textbf{Model} &   \multicolumn{7}{c}{\textbf{Frontal Lobe volume (fro) MAE} $\downarrow$}    \\ 
\cline{2-8} 
        &  \textit{do}(fro)  & \textit{do}(par)  &\textit{do}(tem)&\textit{do}(occ) 
        &\textit{do}(cin)	& \textit{do}(ins)&\textit{do}(ven)                          \\
\hline
VAE   &-& 0.318   & 0.299 & 0.599& 0.725&0.770&\textbf{0.624}  \\
HVAE    &-&\textbf{0.303} &0.292  &0.591 & 0.722&0.769&0.638 \\
VAE-GLM &-& 0.310   & 0.299 &0.594 &0.723 &0.769&0.639 \\
GAN      &-& 0.318& 0.318 &0.602 &0.727 &0.771&0.669 \\
GAN-Finetuned &-&   0.385 & 0.350 &0.620 &0.737 &0.776&0.632\\
HA-GAN &-& 0.311   &\textbf{0.268}  &\textbf{0.588} &\textbf{0.719} &\textbf{0.766}&0.664 \\

\hline
\end{tabular}}
\resizebox{10cm}{!}{
\begin{tabular}{c|ccccccc}
  &   \multicolumn{7}{c}{\textbf{Parietal Lobe volume (par) MAE} $\downarrow$}    \\ 
\cline{2-8}  
        &  \textit{do}(fro)  & \textit{do}(par)  &\textit{do}(tem)&\textit{do}(occ) 
        &\textit{do}(cin)	& \textit{do}(ins)&\textit{do}(ven)                                 \\
\hline
VAE   &   0.423  &-& 0.052 &0.462 &0.654 &0.722&\textbf{0.499} \\
HVAE  &   0.445 &-& \textbf{0.031} &0.449 &0.649 &0.721&0.520 \\
VAE-GLM & 0.433 &-&0.035  &0.453 & 0.651&0.721&0.526 \\
GAN   &  0.385   &-& 0.047 &0.465 & 0.657&0.724&0.570 \\
GAN-Finetuned &  \textbf{ 0.282}  &-&0.084  &0.493 &0.671 &0.731&0.515\\
HA-GAN &   0.475  &-& 0.051 &\textbf{0.445} & \textbf{0.645}&\textbf{0.717}&0.560\\

\hline
\end{tabular}}
\resizebox{10cm}{!}{
\begin{tabular}{c|ccccccc}
  &   \multicolumn{7}{c}{\textbf{Temporal Lobe volume (tem) MAE} $\downarrow$}    \\ 
\cline{2-8} 
        &  \textit{do}(fro)  & \textit{do}(par)  &\textit{do}(tem)&\textit{do}(occ) 
        &\textit{do}(cin)	& \textit{do}(ins)&\textit{do}(ven)                               \\
\hline
VAE   &   0.328  & 0.048   &-&0.374 &0.526 &0.580&\textbf{0.403} \\
HVAE    &  0.346&\textbf{0.029} &-&0.364 &0.522 &0.579&0.420  \\
VAE-GLM & 0.337  & 0.031 &-& 0.367& 0.524&0.580&0.422 \\
GAN  &  0.299 &  0.037  &-&0.377 &0.529 &0.582& 0.459\\
GAN-Finetuned &  \textbf{0.217} & 0.114   &-& 0.399& 0.540&0.588&0.415  \\
HA-GAN &    0.370 &0.039    &-&\textbf{0.360} & \textbf{0.519}&\textbf{0.576}&0.452 \\

\hline
\end{tabular}}
\resizebox{10cm}{!}{
\begin{tabular}{c|ccccccc}
  &   \multicolumn{7}{c}{\textbf{Occipital Lobe volume (occ) MAE} $\downarrow$}    \\ 
\cline{2-8} 
        &  \textit{do}(fro)  & \textit{do}(par)  &\textit{do}(tem)&\textit{do}(occ) 
        &\textit{do}(cin)	& \textit{do}(ins)&\textit{do}(ven)                                \\
\hline
VAE   & 1.711  &0.817    &0.876  &-& 0.403&0.537&0.210 \\
HVAE    &    1.755 & 0.862 &0.896  &-& 0.394&0.534&\textbf{0.191} \\
VAE-GLM &   1.731 &0.843  & 0.875 &-& 0.398&0.536&0.215 \\
GAN      &  1.637 & 0.819 &0.819  &-& 0.410&0.542&0.313 \\
GAN-Finetuned &   \textbf{1.434}  & \textbf{0.617} & \textbf{0.723} &-&0.438 &0.556&0.270 \\
HA-GAN &   1.814  & 0.839   & 0.967 &-&\textbf{0.386} &\textbf{0.527}&0.220 \\

\hline
\end{tabular}}
\resizebox{10cm}{!}{
\begin{tabular}{c|ccccccc}
  &   \multicolumn{7}{c}{\textbf{Cingulate cortex volume (cin) MAE} $\downarrow$}    \\ 
\cline{2-8} 
        &  \textit{do}(fro)  & \textit{do}(par)  &\textit{do}(tem)&\textit{do}(occ) 
        &\textit{do}(cin)	& \textit{do}(ins)&\textit{do}(ven)                                \\
\hline
VAE   & 4.329  & 2.490   &2.612  & 0.758&- &0.292&0.619\\
HVAE    & 4.418  & 2.584   &2.653  &0.810 &- &0.287&0.521 \\
VAE-GLM & 4.369 & 2.545   &2.610  &0.793 &-&0.290&0.524\\
GAN      &   4.176  & 2.495  &2.494  &0.743 &-&0.303&0.449\\
GAN-Finetuned & \textbf{3.760}    & \textbf{2.079}   &\textbf{2.297}  &\textbf{0.632} &-&0.331&0.610 \\
HA-GAN &  4.540   & 2.536   &2.800  &0.826 &-&\textbf{0.272}&\textbf{0.360} \\

\hline
\end{tabular}}
\resizebox{10cm}{!}{
\begin{tabular}{c|ccccccc}
  &   \multicolumn{7}{c}{\textbf{Insula volume (ins) MAE} $\downarrow$}    \\ 
\cline{2-8} 
        &  \textit{do}(fro)  & \textit{do}(par)  &\textit{do}(tem)&\textit{do}(occ) 
        &\textit{do}(cin)	& \textit{do}(ins)&\textit{do}(ven)                                 \\
\hline
VAE   &  6.930 &4.168 & 4.351 &1.566 &0.403 &-&1.341 \\
HVAE  & 7.064  &4.308    &4.412  &1.645 &0.431 &-&1.210 \\
VAE-GLM &  6.990 & 4.250 & 4.349 &1.618 &0.419 &-&1.197 \\
GAN      & 6.701  &  4.176  &4.174  &1.544 &0.382 &-& \textbf{0.936}\\
GAN-Finetuned & \textbf{6.075}  &\textbf{3.550}    &\textbf{3.878}  &\textbf{1.378} &\textbf{0.294} &-&1.264 \\
HA-GAN &    7.246 &4.236 &4.634  &1.669 &0.455 &-&0.968 \\

\hline
\end{tabular}}
\resizebox{10cm}{!}{
\begin{tabular}{c|ccccccc}
  &   \multicolumn{7}{c}{\textbf{Ventricle volume (ven) MAE} $\downarrow$}    \\ 
\cline{2-8} 
        &  \textit{do}(fro)  & \textit{do}(par)  &\textit{do}(tem)&\textit{do}(occ) 
        &\textit{do}(cin)	& \textit{do}(ins)&\textit{do}(ven)                              \\
\hline
VAE   &  0.827 & 0.426   & 0.451 &0.112 &0.134 &0.188&-\\
HVAE    & 0.846 &0.447  &0.461  &0.120 & 0.134&0.187& -\\
VAE-GLM & 0.836  & 0.438   &0.451  &0.119 & 0.135&0.187& -\\
GAN      &0.793 &  0.428  &0.426  &0.117 &0.138 &0.190&-\\
GAN-Finetuned &  \textbf{0.702} & \textbf{0.340}   &\textbf{0.384}  & \textbf{0.109}&0.147 &0.196&-\\
HA-GAN &   0.873  &0.435    &0.492  &0.117 &\textbf{0.130} &\textbf{0.183}& -\\

\hline
\end{tabular}}
\end{table}

\end{document}